Laser-induced plasma generation of terahertz radiation using three incommensurate wavelengths

Jacob D. Bagley[a], Clayton D. Moss[a], and Jeremy A. Johnson[b]

Department of Chemistry and Biochemistry, Brigham Young University, Provo, UT, 84602, USA

**Abstract**

We present the generation of THz radiation by focusing ultrafast laser pulses with three incommensurate wavelengths to form a plasma. The three colors include 800 nm and the variable IR signal and idler outputs from an optical parametric amplifier. Stable THz is generated when all three colors are present, with a peak-to-peak field strength of ~200 kV/cm and a relatively broad, smooth spectrum extending out to 6 THz, without any strong dependence on the selection of signal and idler IR wavelengths (in the range from 1300-2000 nm). We confirm that 3 colors are indeed needed, and comment on the polarization characteristics of the generated THz, some of which are challenging to explain with plasma current models that have had success in describing two-color plasma THz generation.

[a] Contributed equally to this work.
[b] Electronic Mail: jjohnson@chem.byu.edu

**Introduction**

Ultrafast spectroscopy utilizing high-field terahertz (THz) radiation (> 0.1 MV/cm) has proven to be a valuable tool in studying many intriguing phenomena. Several material properties couple to degrees of freedom, such as rotations and collective vibrations, that interact with THz radiation. Therefore, powerful applications are utilizing THz in new and exciting ways, including chemical recognition [1], security and imaging applications [2-5], industrial-scale, non-destructive monitoring systems in industry [6] and food processing [4], studying nuanced interactions in biological systems [7], extremely high-bandwidth wireless communication, and developing future generation high-speed electronic and computational devices [8]. Cutting edge fundamental studies are also made possible with *intense* THz pulses, including the first direct observation of an electromagnon [9] and the coherent control of collective modes in gases, liquids and solids [10-15].

These experiments are made possible in part due to important advancements in the generation of strong THz pulses. Favored methods of THz generation include photoconductive switches [16-18], optical rectification in nonlinear-optical crystals [19,20], and two-color laser plasma generation [21-23]. Thus far, optical rectification and laser-induced plasma generation of THz are able to produce the pulses with highest peak electric-field strengths. Each method has advantages and disadvantages; laser-induced plasma generation advantageously results in spectrally broad pulses that lack absorptive features inherent to THz generation crystals [24]. The typical two-color laser plasma generation scheme produces terahertz pulses by focusing two separate laser pulses of commensurate wavelengths, e.g., a fundamental wavelength λ overlapped with its second harmonic λ/2, with the second-harmonic often generated by focusing through a nonlinear optical crystal like BBO. The focused light has high enough intensity to create a plasma filament, and the two-color electric field is understood to create an asymmetrical electrical current in the plasma that emits broadband THz radiation [25-28]. This scheme can generate continuous bandwidth radiation ranging from 0.1 THz to 5–200 THz, depending on the duration of the pump pulse [29,30] with electric field strengths reported up to several MV/cm [21]. Despite the potential utility of plasma-based THz generation schemes and recent advances in understanding the mechanism of THz plasma generation, it is an intense area of research [21,27,31-34] with a number of theoretical details still not completely understood.

Typically, two-color plasma generation of THz is carried out using commensurate wavelengths (i.e. a fundamental wavelength and its second harmonic). Some attempts have been made to extend this to incommensurate wavelengths where the wavelength of one color is not an integer multiple of the other. Using incommensurate wavelengths results in a relative phase between the two electric fields that depends on time, potentially leading to high-frequency modulations of the total electric field envelope, the resulting photocurrent, and the generated THz waveform [31]. Vvedenskii *et al*. used the tunable output of an optical parametric amplifier (OPA) in an initial test of mixing incommensurate wavelengths [32]. Starting with commensurate wavelengths ($\lambda = 1600$ nm and $\lambda/2 = 800$ nm), they monitored THz generation while detuning the longer wavelength (1600 nm, the signal wavelength derived from the OPA). As shown in Figure 2(b) below (red circles and accompanying line), varying the longer wavelength from 1600 nm down to ~1500 nm and mixing with 800 nm resulted in complete loss of THz output. Later, this work was theoretically extended to cover a larger range of wavelengths and predict THz yields by mixing two colors with a variety of frequency ratios [35]. Performing experiments with longer incommensurate wavelengths, Balčiūnas *et al*. found they could extend the range of detuned-THz generation indefinitely, but showed that a carrier-envelope-phase (CEP) stabilized laser was necessary [31]; turning off the laser CEP stabilization disrupted the production of CEP-stable THz pulses. Additionally, Balčiūnas *et al*. found that the high-frequency modulations to the total electric field envelope inherent when using incommensurate wavelengths led to additional narrowband, high-frequency (>10 THz) components of the generated THz [31].

Here we demonstrate the generation of stable THz pulses using a very broad range of incommensurate wavelengths, but without a CEP-stable laser system. This is accomplished by adding a third incommensurate wavelength to the plasma. In our experiments, an OPA was adjusted such that three incommensurate wavelengths (800 nm and two IR colors corresponding to the signal and idler) were output collinearly and subsequently focused together to generate the laser-induced plasma. Varying the wavelengths of the signal and idler beams, we observe nearly identical THz generation over a broad range of wavelengths. We also investigate the polarization dependence of the incident beams and how it influences the THz polarization. Some of the results are challenging to explain with the current theoretical understanding of laser-induced plasma THz generation.

**Methods**

An amplified Ti:Sapphire laser system is used to generate ~100 fs laser pulses with 800-nm central wavelength. A beam splitter divides the pump beam and the probe beam (used for electro-optic sampling of the THz pulses). The pump pulse is directed to an OPA for down-conversion to IR (signal and idler) wavelengths. In a standard two-stage, white-light seeded OPA design, a small part of the 800-nm pump creates a seed IR beam at the first stage, which is amplified by the remainder of the pump at a second stage. As shown in Figure 1(b), our OPA is reconfigurable to perform $2^{nd}$ stage pumping with a portion of the $2^{nd}$-stage pump beam, allowing the other portion of 800-nm light to be recombined and co-propagate out of the OPA with the signal and idler beams (commonly called a "fresh-pump" configuration, typically used for sum-frequency generation). This allows us to collinearly combine the three colors (800 nm + signal + idler) in time, and then use an off-axis parabolic mirror to focus them together making a plasma for THz generation. Before recombining 800 nm with signal and idler, a delay stage is used to overlap the 800-nm pulse in time with the signal and idler pulses as shown in Fig. 1(b), and a $\lambda/2$ waveplate is used for polarization control.

The signal wavelength was varied from 1600 to 1300 nm with the corresponding idler wavelength ranging from ~1600 to 2075 nm with the signal polarization vertical and the idler polarization horizontal relative to the laboratory frame. The polarization of the co-propagating 800-nm beam could be rotated by a $\lambda/2$ waveplate to arbitrary polarizations (see Fig. 1(b)); the polarization was set to either be parallel to the signal or idler polarization, or at 45° in between the two. As shown in Fig. 1(a), the three pulses were focused to a plasma using an off-axis parabolic mirror. A thin sheet of Teflon placed after the plasma acted as low pass filter removing the incident signal, idler, 800 nm, and other generated colors, while passing the generated THz radiation. Two additional off-axis parabolic mirrors were used to collimate and then refocus the THz to a 100 μm (110) GaP bonded to a 1 mm (100) GaP crystal for electro-optic sampling with 800-nm probe pulses. Two main sets of measurements were performed: (1) Signal and idler polarizations were kept perpendicular and the 800-nm polarization was rotated to be parallel with either signal or idler, and we varied the signal and idler wavelengths (see Table 1). Although signal and idler powers varied with wavelength, the pulse energies were ~0.40 mJ, ~0.32 mJ, and 0.8 mJ for signal, idler, and 800 nm respectively. (2) A calcite Glan-laser polarizer was placed at the output of the OPA to select a uniform polarization for all three signal, idler, and 800 nm beams. As the polarizer was rotated for different measurements, at some polarizer angles one of the three beams was completely suppressed, allowing us to check the necessity of all three-colors being present. For the second set of measurements, the signal and idler wavelengths were fixed at 1550 nm and 1653 nm respectively.

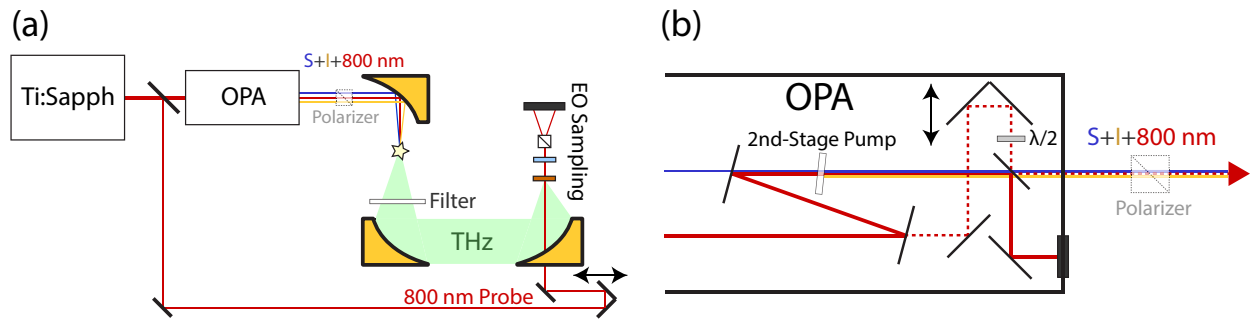

**Figure 1.** (color online) (a) Illustration of the experimental setup. The original 800-nm light is split into a probe beam and a pump beam for the OPA. From the OPA a "fresh" 800-nm pump, the signal (S) and idler (I) pulses are focused to form a plasma using an off-axis parabolic mirror. A Teflon filter passes the generated THz radiation, and the THz waveform is measured using electro-optic (EO) sampling. A polarizer is placed in the S+I+800-nm beam path in the second set of measurements. (b) Depiction of the final OPA beam paths including the $2^{nd}$-stage pump. The dotted line shows the path of the "fresh" 800-nm pulse, with variable time delay and polarization control, which leaves collinearly with the amplified signal and idler pulses.

We also point out that relative-phase stability is required between all 3-colors in both sets of measurements to generate THz. To achieve temporal overlap between the 800-nm pulses and the two IR colors, we adjust the delay stage in the OPA shown in Fig. 1(b). However, the smallest possible changes in the stage position lead to wild jumps in the THz output (for a given probe delay) because the stage motion is not accurate enough to reproduce positions with optimal relative phases. Thus, we used this delay stage for rough temporal alignment, and for fine-timing adjustment we slightly rotated the $2^{nd}$-stage pump nonlinear crystal in the OPA. Such crystal rotation is typically done to optimize the phase-matching conditions between the 800-nm pump and the seed-IR light, but it also results in a very fine phase-delay adjustment for the generated signal and idler pulses that allows finding the optimal phase delay, relative to the 800-nm, in the plasma. The range of motion for this fine-timing adjustment was not enough to significantly modify the IR output.

**Results and Discussion**

Figure 2(a) displays representative 3-color THz traces of data set (1), where the Signal (S) or Idler (I) wavelength is indicated and the inset depicts the corresponding Fourier transforms with a broad spectrum extending out to 6 THz and peak-to-peak electric field strengths of ~200 kV/cm. In these measurements, we varied the signal and idler wavelengths and extracted the THz peak-to-peak signals for each set of wavelengths, as shown in Figure 2(b). In all measurements, all three wavelengths were present. For wavelengths below 1600 nm (blue triangles) the markers indicate the signal wavelength (vertically polarized), the 800 nm was vertically polarized parallel to the signal polarization, and the idler was horizontal; interestingly, the THz polarization was horizontal (determined by recording two measurements with the GaP EO crystal oriented to detect either vertical or horizontal THz). For measurements above 1600 nm (the yellow-orange squares) the markers indicate the idler wavelength (horizontally polarized); in these measurements, the 800-nm pulse is horizontally polarized parallel to the idler, and the THz polarization was vertical, parallel to the signal polarization. In this configuration, the emitted THz polarization is always *perpendicular* to the polarization of the 800 nm + signal (idler), and *parallel* to the third color: the idler (signal), as summarized in Table 1.

Table 1. Set wavelengths and polarizations for data set (1) depicted in Fig. 2.

|  | < 1600 nm | > 1600 nm |
| --- | --- | --- |
| Set Wavelength (Pol) | Signal (V) | Idler (H) |
| 800-nm Polarization | V | H |
| THz Polarization | H | V |

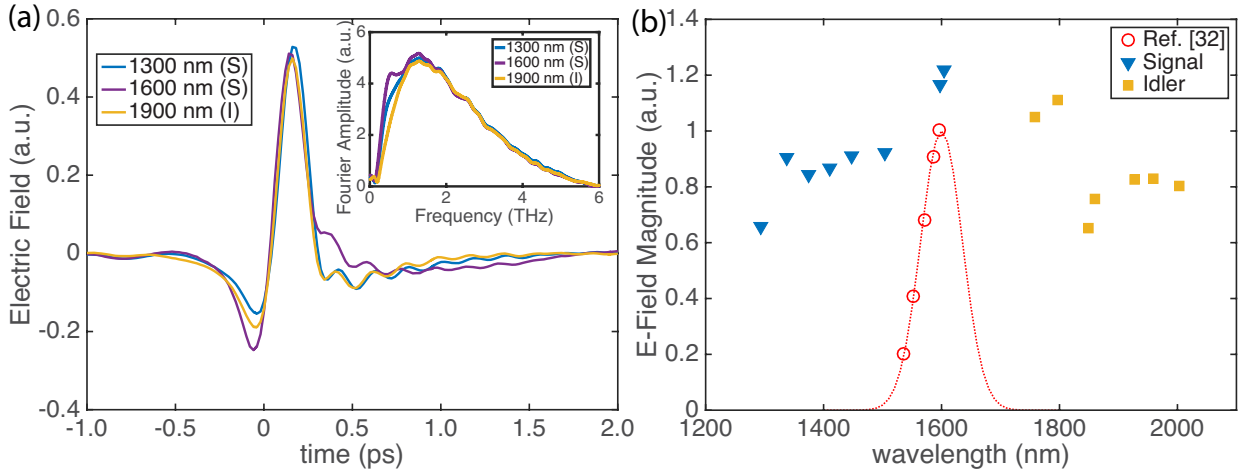

**Figure 2.** (color online) (a) Representative THz waveforms recorded at three set-wavelengths. The inset shows the respective Fourier transforms. (b) THz peak-to-peak electric field strength generated from a plasma filament as a function of signal wavelength (blue triangles) and idler wavelength (yellow-orange squares) as described in the text. The red circles and curve represent the two-incommensurate wavelength results from Ref. [32].

When mixing three incommensurate colors, we do not observe a reduction in THz output as the signal and idler wavelengths are detuned from the commensurate 1600 nm as observed in Ref [32] when mixing two incommensurate colors. In the range of 1300–1800 nm, the THz yield is fairly uniform, and although the THz yield decreases at wavelengths greater than 1800 nm and approaching 1300 nm, in these regions our OPA outputs less power. Using neither commensurate wavelengths nor a CEP stable laser system [31,32], we show that THz output is nearly identical across this very broad range of wavelengths when a third wave component is present. To test this assertion that all three wavelengths are needed, we performed a second set of measurements, as described above, where a broadband polarizer was inserted into the common path, making the polarization of all three transmitted wavelengths uniform.

Figure 3 displays the results of using the polarizer to control the amount of signal, idler, and 800 nm beams that were incident on the plasma filament during THz generation. In these measurements, the signal and idler wavelengths were fixed at 1550 nm and 1653 nm with initial vertical and horizontal polarizations, respectively. For Fig. 3 panels (a), (b), and (c) the 800-nm beam was initially polarized horizontally, and for panels (d), (e), and (f) the 800-nm beam was polarized 45° from horizontal. Panels (a) and (d) show the calculated fraction of the 800 nm (red-orange), signal (1550 nm, yellow-orange), and idler (1653 nm, dashed blue) beams that transmit through the polarizer at each angle. Panels (b) and (e) and panels (c) and (f) display detecting horizontal and vertical THz components respectively, recorded by rotating the EO crystal 90 degrees between sets of measurements. The triangles in panels (b), (c), (e), and (f) display the measured terahertz peak-to-peak electric field strength at each polarizer angle. Finally, the purple dashed line in panels (b), (c), (e), and (f) is the mathematical product of the three beam intensities (3-wave-product) in (a) and (d), providing an indication of when one wavelength is completely suppressed due to the polarization selection.

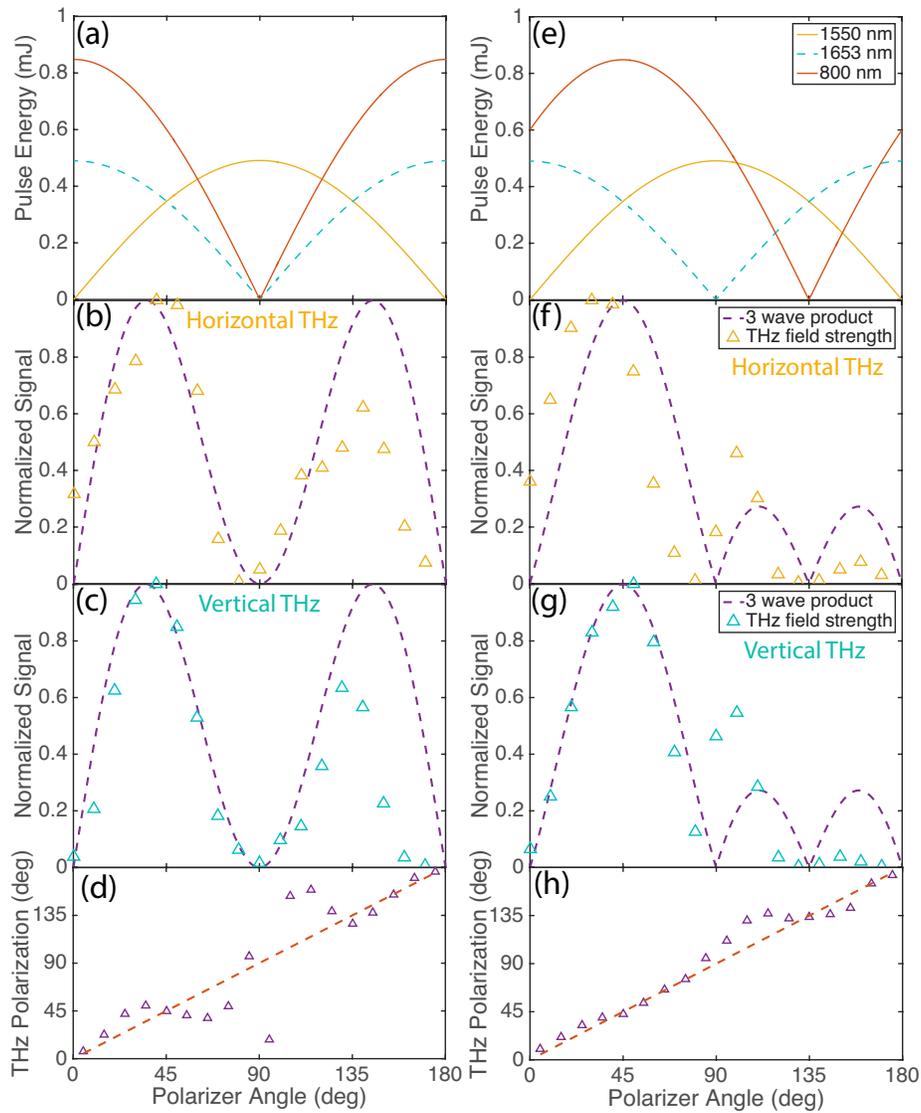

**Figure 3.** (color online) Incident pulse energies (a,e) and generated THz peak-to-peak field strengths (b-c,f-g) as a function of polarizer angle. A rough calculation of the THz polarization is shown in (d,h). In all plots, 1550 nm is initially vertical polarization (90 degrees) and 1653 nm is initially horizontal polarization (0 degrees). In (a-d) the 800-nm light is initially horizontally polarized (0 degrees) and in (e-h), the 800-nm light is initially polarized at 45 degrees. (a) and (e) show incident pulse energies as a function of polarizer angle. (b) and (f) show the peak-to-peak THz field strengths when the EO crystal is oriented to detect horizontally polarized THz pulses, and for the same setup, (c) and (g) show the peak-to-peak THz field strengths when the EO crystal is rotated to detect vertically polarized THz pulses. The dashed purple lines in (b-c,f-g) indicate the product of the three pulse energies displayed in (a) and (e).

When only two beams are present in the plasma filament (e.g. polarizer angles of 0°, 90°, or 180° for Fig. 3 panels (a), (b), (c) and polarizer angles of 0°, 90°, 135°, or 180° for panels (d), (e), (f)) THz is not produced. However, stable THz is always produced when a third beam is present. The 3-wave-products roughly demonstrates that maximizing all three beams produces the most THz. Some discrepancies between the 3-wave-product and the THz field strength are seen at larger polarizer angles; we attribute this discrepancy to potential spatial and temporal walk-off between the three beams due to small alignment changes as the polarizer is rotated and group-velocity mismatch between the three colors in the polarizer. The plasma THz-generation process is also highly nonlinear and peak-to-peak field strengths do not vary linearly with intensity. In any case, two incommensurate wave components do not create stable THz, but three wavelength components together are necessary to generate THz radiation from the plasma filament in this experimental apparatus.

In Figure 3(d) and 3(h) we show a rough determination of the THz polarization, showing that when all three colors have uniform polarization, the generated THz maintains nearly the same polarization. Figure 4 shows recorded THz waveforms with the GaP EO-crystal angle set to record either horizontally polarized (H) or vertically polarized (V) THz pulses. Figure 4(a) displays the H and V THz components for two different polarizer angles when corresponding to H 800-nm light as in Fig. 3(a-d), and Fig. 4(b) shows THz traces when the 800-nm light is polarized at 45° as in Fig. 3(e-g). The THz polarization reported in Fig. 3(d,h) was obtained by taking the arctangent of the ratio of electric field amplitudes from traces like those in Fig. 4; small THz signals leads to some of the scatter seen in Fig. 3(d,h).

We will comment on two important features in Fig. 4. First, we see that the horizontally polarized THz changes sign when rotating the polarizer from 40° to 140°, whereas the vertically polarized THz maintains the same sign. This indicates that the total THz polarization has rotated from +40° to +140° as the pump polarizer angle was rotated. Secondly, we note that the group delay of the horizontal THz is slightly behind that of the vertical THz, indicating that there is some amount of ellipticity to the overall THz polarization; this could be due to small changes in relative group-delay between the 3-colors. Furthermore, the waveforms are different between H and V THz resulting in differences in the corresponding spectra and a somewhat complicated total polarization and waveform. Similar features are seen in Fig. 4(b) as well. Thus we emphasize that the THz polarization shown in Fig. 3(d,h) is a rough approximation, determined with the peak-field strengths of the H and V THz components with no respect for relative group delay. Future measurements with a experimental setup where the polarization and relative delay of the three pulses could be changed independently would be useful in exploring the prospect of frequency-specific polarization control.

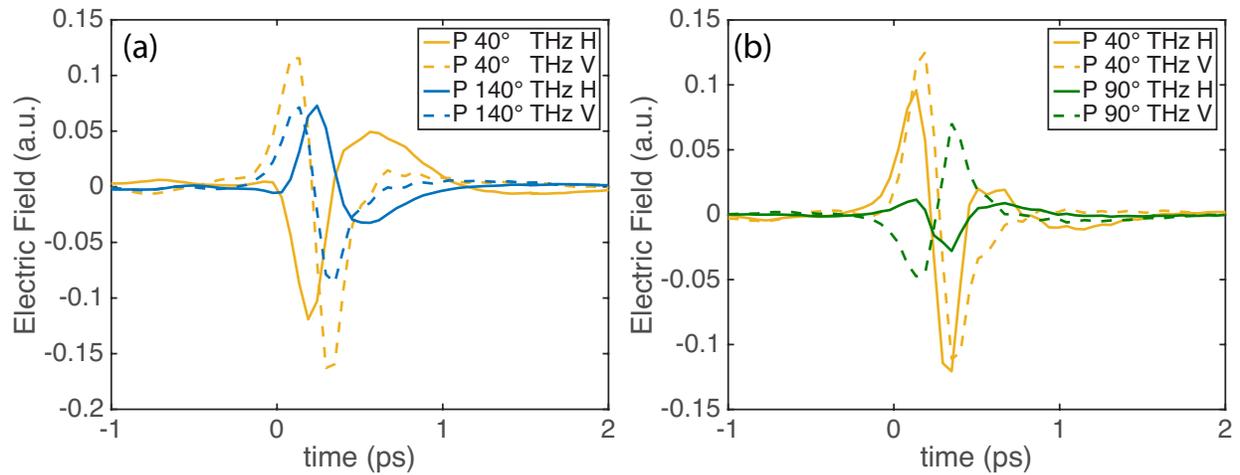

**Figure 4.** (color online) THz traces recorded to recover both horizontal (H – solid lines) and vertical (V – dashed lines) components. (a) Measurements with the 800-nm light initially polarized horizontally. The yellow-orange lines were recorded with the pump-polarizer set to 40°; the blue lines correspond to 140°. (b) Measurements with the 800-nm light initially polarized at 45°. The yellow-orange lines were recorded with the pump-polarizer set to 40°; the green lines correspond to 90°.

We conclude with a discussion of certain aspects of these results that are challenging to explain within the framework of plasma current models. In commensurate-two-color plasma experiments, it is generally understood that fundamental and second-harmonic wavelengths of the same polarization mix to asymmetrically accelerate charge carriers and build up the plasma current that is responsible for THz radiation (with the same polarization as the mixed pump colors) [26,27,35,36]. This idea matches well with the second set of measurements where the polarizer forces all three pump colors to have the same polarization, and the resulting THz radiation has basically the same polarization as the incident light. But in the first set of measurements where two colors have the same polarization, and the third is perpendicular, we observe that the THz radiation always has the same polarization as the third color. One color on its own should not be able to generate an asymmetric current, and yet we observe THz generation with identical polarization in these conditions. Further measurements with independent timing, polarization, and power control of the three beams would be ideal to better understand the mechanism for stable THz generation.

**Conclusion**

In summary, we have presented a scheme for generating stable THz pulses using ultrafast pulses with three incommensurate wavelengths. The THz pulses have smooth and broad spectra extending out to 6 THz. In contrast to previous measurements needing a CEP-stabilized laser system for THz with incommensurate wavelengths, adding a third wavelength stabilizes the generation of CEP-stable THz pulses. We showed that all three colors are needed for stable THz generation and reported on the THz polarization characteristics in two main sets of measurements. The current results are challenging to explain with plasma current models that have had much success in describing two-color plasma THz generation with commensurate and incommensurate wavelengths [21,26,27,31,32,35,36].

Although challenging due to the relative phase-stability requirements, future work allowing individual polarization, delay, and power control over the three incommensurate colors could prove instructive and useful in testing the THz generation efficiency and the impact on the generated THz polarization, potentially allowing THz waveforms with custom polarization states and higher electric-field strengths. Furthermore, measurements with higher bandwidth than our

~100 fs probe pulses allowed could check the potential for tunable (as the signal and idler wavelengths are varied [31]), narrow-band THz generation at higher frequencies than we could measure here.

**Acknowledgments**

The authors acknowledge funding and support from the Department of Chemistry and Biochemistry and the College of Physical and Mathematical Sciences at Brigham Young University.

**References**

[1]     J. B. Baxter and G. W. Guglietta, Analytical Chemistry **83**, 4342 (2011).
[2]     C. Wai Lam, D. Jason, and M. M. Daniel, Reports on Progress in Physics **70**, 1325 (2007).
[3]     A. Redo-Sanchez, B. Heshmat, A. Aghasi, S. Naqvi, M. Zhang, J. Romberg, and R. Raskar, Nature Communications **7**, 12665 (2016).
[4]     A. A. Gowen, C. O'Sullivan, and C. P. O'Donnell, Trends in Food Science & Technology **25**, 40 (2012).
[5]     D. M. Mittleman, R. H. Jacobsen, and M. C. Nuss, IEEE Journal of Selected Topics in Quantum Electronics **2**, 679 (1996).
[6]     W. Zouaghi, M. D. Thomson, K. Rabia, R. Hahn, V. Blank, and H. G. Roskos, European Journal of Physics **34**, S179 (2013).
[7]     K. Shiraga, Y. Ogawa, T. Suzuki, N. Kondo, A. Irisawa, and M. Imamura, Applied Physics Letters **102**, 053702 (2013).
[8]     I. F. Akyildiz, J. M. Jornet, and C. Han, Physical Communication **12**, 16 (2014).
[9]     T. Kubacka *et al.*, Science **343**, 1333 (2014).
[10]    T. Huber, M. Ranke, A. Ferrer, L. Huber, and S. L. Johnson, Applied Physics Letters **107**, 091107 (2015).
[11]    H. Y. Hwang *et al.*, Journal of Modern Optics **62**, 1447 (2015).
[12]    T. Kampfrath, K. Tanaka, and K. A. Nelson, Nat Photon **7**, 680 (2013).
[13]    I. Katayama *et al.*, Physical Review Letters **108**, 097401 (2012).
[14]    S. Fleischer, R. W. Field, and K. A. Nelson, Physical Review Letters **109**, 123603 (2012).
[15]    J. Lu, Y. Zhang, H. Y. Hwang, B. K. Ofori-Okai, S. Fleischer, and K. A. Nelson, Proceedings of the National Academy of Sciences **113**, 11800 (2016).
[16]    D. H. Auston, K. P. Cheung, and P. R. Smith, Applied Physics Letters **45**, 284 (1984).
[17]    C. Fattinger and D. Grischkowsky, Applied Physics Letters **53**, 1480 (1988).
[18]    P. U. Jepsen, R. H. Jacobsen, and S. R. Keiding, J. Opt. Soc. Am. B **13**, 2424 (1996).
[19]    C. Vicario, M. Jazbinsek, A. V. Ovchinnikov, O. V. Chefonov, S. I. Ashitkov, M. B. Agranat, and C. P. Hauri, Opt. Express **23**, 4573 (2015).
[20]    S.-H. Lee, S.-J. Lee, M. Jazbinsek, B. J. Kang, F. Rotermund, and O. P. Kwon, CrystEngComm **18**, 7311 (2016).
[21]    M. Clerici *et al.*, Physical Review Letters **110**, 253901 (2013).
[22]    D. J. Cook and R. M. Hochstrasser, Opt. Lett. **25**, 1210 (2000).

[23] T. Bartel, P. Gaal, K. Reimann, M. Woerner, and T. Elsaesser, Opt. Lett. **30**, 2805 (2005).
[24] K. Y. Kim, J. H. Glownia, A. J. Taylor, and G. Rodriguez, IEEE Journal of Quantum Electronics **48**, 797 (2012).
[25] V. A. Andreeva *et al.*, Physical Review Letters **116**, 063902 (2016).
[26] K.-Y. Kim, Physics of Plasmas **16**, 056706 (2009).
[27] K. Y. Kim, J. H. Glownia, A. J. Taylor, and G. Rodriguez, Opt. Express **15**, 4577 (2007).
[28] M. Li, W. Li, Y. Shi, P. Lu, H. Pan, and H. Zeng, Applied Physics Letters **101**, 161104 (2012).
[29] V. Blank, M. D. Thomson, and H. G. Roskos, New Journal of Physics **15**, 075023 (2013).
[30] E. Matsubara, M. Nagai, and M. Ashida, Applied Physics Letters **101**, 011105 (2012).
[31] T. Balčiūnas *et al.*, Opt. Express **23**, 15278 (2015).
[32] N. V. Vvedenskii, A. I. Korytin, V. A. Kostin, A. A. Murzanev, A. A. Silaev, and A. N. Stepanov, Physical Review Letters **112**, 055004 (2014).
[33] K. Pernille, C. S. Andrew, I. Krzysztof, and J. Peter Uhd, New Journal of Physics **15**, 075012 (2013).
[34] X. Xie, J. Dai, and X. C. Zhang, Physical Review Letters **96**, 075005 (2006).
[35] V. A. Kostin, I. D. Laryushin, A. A. Silaev, and N. V. Vvedenskii, Physical Review Letters **117**, 035003 (2016).
[36] I. Babushkin, W. Kuehn, C. Köhler, S. Skupin, L. Bergé, K. Reimann, M. Woerner, J. Herrmann, and T. Elsaesser, Physical Review Letters **105**, 053903 (2010).